\documentstyle[12pt,emulateapj]{article}

\lefthead{Yadav et al.}
\righthead{Different types of X-ray bursts}

\slugcomment{To appear in June 1, 1999 issue of The Astrophysical Journal}

\begin{document}

\title{Different types of X-ray bursts from  GRS~1915+105 \\
       and  their origin}

\author{
	J.~S.~Yadav\altaffilmark{1},
	 A.~R.~Rao,
	 P.~C.~Agrawal,
	 B.~Paul}
\affil{Tata Institute of Fundamental Research, Homi Bhabha Road, Mumbai 400 005, India}

\altaffiltext{1}{Send offprint request to: jsyadav@tifr.res.in}

\and

\author{S. Seetha, and K. Kasturirangan}
\affil{ISRO Satellite Centre, Airport Road, Vimanpura P.O. Bangalore  560 017, India}

\begin{abstract}
We report the X-ray observations of the Galactic  X-ray transient 
source GRS~1915+105 with the Pointed Proportional Counters of the Indian 
X-ray Astronomy Experiment(IXAE) onboard the Indian satellite IRS-P3, 
which show remarkable richness in temporal variability.  The observations 
were carried out on 1997 June 12 - 29 and  August 7 - 10, in the energy 
range of 2$-$18 keV and revealed the presence of  very intense X-ray 
bursts.   All the observed bursts have a slow exponential  rise, a sharp  
linear decay, and they    can  broadly be put in two classes: irregular 
and quasi-regular bursts in one class, and  regular bursts in another 
class. The regular bursts are found to have  two distinct time scales 
and they persist over extended durations.  
There is a strong correlation between the preceding quiescent time and 
the burst duration for the quasi-regular and irregular bursts. No such 
correlation is found for the regular bursts.   The ratio of average flux 
during the burst time to the average flux during the quiescent phase is 
high and variable  for the quasi-regular and irregular bursts while it 
is low and constant for the regular bursts.
We present a  comprehensive picture of the various types of bursts
observed in GRS 1915+105 in the light of the recent theories
of advective accretion disks.  We suggest that the peculiar bursts that 
we have seen are characteristic of the change of state of the source. 
The source can switch back and forth between the low-hard state and
the high-soft state  near  critical accretion rates in a very short time 
scale, giving rise to the irregular and quasi-regular  bursts.  The fast 
time scale for the transition of the state is explained by invoking the 
appearance and disappearance of the advective disk in its viscous time 
scale.   The periodicity of the regular bursts is explained by matching  
the viscous time scale with the cooling time scale of the post shock 
region.  A test of the model is presented using the publicly available 
13$-$60 keV RXTE/PCA data for irregular and regular bursts concurrent 
with our observations. It is found that the 13$-$60 keV flux relative to 
the 2$-$13 keV flux shows clear evidence for state change between the 
quiescent phase and the burst phase. The value of this ratio during 
burst is consistent with the values observed during the high-shoft 
state seen on 1997 August 19 while its value during quiescent phase is 
consistent with the values observed during the low-hard state seen 
on 1997 May 8. 
\end{abstract}

\section{Introduction}

The X-ray transient source GRS~1915+105 was discovered in 1992 with the WATCH 
all sky X-ray monitor onboard the GRANAT satellite (\cite{cast:94}).   
Superluminal motions of two symmetric radio emitting jets of GRS~1915+105 
were discovered by Mirabel \& Rodriguez (\cite{mira:94}).   Several 
features in the observed properties of GRS~1915+105 such as the Power 
Density Spectra (PDS) with the QPO feature, a hard X-ray tail and 
the subsecond time variability, are typical characteristics of black 
hole binaries. The X-ray intensity is found to be more than
$10^{39}$ erg s$^{-1}$ (based on an assumed distance of 12.5 kpc) for 
extended periods which is super-Eddington
luminosity for a neutron star (\cite{mira:94}). 
The other Galactic source of superluminal jets, GRO~J1655-40, has been shown
to harbor a compact object of mass $\sim$7 M$_\odot$ (\cite{oros:97}). The 
combination of relativistic jets and a central black hole has earned these 
two objects the name ``microquasars'' as they seem to be stellar mass analogs 
of the massive black hole systems in quasars and other active galactic nuclei (AGNs).
These microquasars have opened the possibility of studying phenomena in our
Galaxy that until recently were believed  to be restricted to distant quasars
and a few AGNs. In particular, it has been realized that since the 
characteristic dynamic times in the flow of matter onto  a black hole are
proportional to its mass, the events with intervals of minutes in a microquasar
could correspond to analogous phenomena with  duration of thousands of years in 
a quasar of 10$^9$ M$_\odot$.

GRS 1915+105 was observed to be X-ray active in 1994 using BATSE instrument. 
The source went into a very high X-ray 
intensity state in early 1996 and 
was observed on several occasions by the Pointed Proportional Counters (PPCs) 
of the Indian X-ray Astronomy Experiment (IXAE) (\cite{agra:97}; 
\cite{paul:97}), the Proportional Counter Array (PCA) and the All 
Sky Monitor (ASM) of the Rossi X-ray Timing Explorer (RXTE) (\cite{brad:96}).
The X-ray intensity was found to vary on a variety of time scales and
the light curve showed a complicated pattern of dips and rapid transitions 
between high and low intensity (\cite{grei:96}; \cite{bell:97b}; 
\cite{taam:97}).  PPC observations of GRS~1915+105 in its low hard 
state in 1996 July
showed intensity variations by a factor of 2 to 3 at 100$-$400 ms time
scale (\cite{paul:97}; \cite{paul:98a}). Strong (rms variability 9\%) and narrow
(${\nu \over \delta \nu} \approx 5$) Quasi Periodic Oscillations (QPOs) of
varying frequency were discovered in GRS~1915+105 with the PPC
observations (\cite{agra:96}). Quasi-regular X-ray and infrared (IR) flares
with a spacing of $\sim$30 minutes, were observed during simultaneous X-ray/IR 
observations (\cite{eike:98}). These observations suggest that IR flares are
signatures of plasma ejection in the inner part of the accretion disk
which are termed as ``baby jets'' analog to the much larger superluminal
ejection events. At later
times, the X-ray flares decouple from IR flares ruling out thermal reprocessing of the X-rays as the source of the IR flares. Another simultaneous observations
of GRS 1915+105 in the X-ray, IR, and radio wavelengths confirm that the IR and radio flares are associated with the X-ray dips (\cite{mira:98}). 


\begin{deluxetable}{clllcccc}
\footnotesize 
\tablecaption{The  GRS~1915+105 observations  by IXAE during 1997 June-August.   \label{tbl-1}}
\tablewidth{0pt}
\tablehead{
\colhead{DOY} & \colhead{Date} &   \colhead{Start time} & \colhead{End time} & \colhead{Orbits}&\colhead{time(s)}& \colhead{Mode} &  \colhead{PPC} 
}
\startdata
 164& Jun 12 & 17:36:57 &  19:27:24  &  2 &  1600 & M  &  1 2 3\nl
 168& Jun 16 & 14:21:46 &  18:04:10  &  4 &  1880 & M  &  1 2 3\nl
 169& Jun 17 & 13:59:54 &  17:42:50  &  3 &  2760 & M  &  - 2 3\nl
 169& Jun 18 & 11:55:40 &  15:40:19  &  3 &  1630 & M  &  - 2 3\nl
 170& Jun 19 & 11:35:54 &  15:20:26  &  3 &  1530 & M  &  1 2 3\nl
 173& Jun 21 & 10:59:28 &  17:55:38  &  5 &  5070 & N  &  1 2 3\nl
 174& Jun 22 & 12:12:24 &  19:20:50  &  5 &  5300 & N  &  1 2 3\nl
 175& Jun 23 & 11:52:08 &  18:59:46  &  5 &  5400 & N  &  1 2 3\nl
 176& Jun 24 & 11:30:15 &  18:38:52  &  5 &  5700 & N  &  1 2 3\nl
 177& Jun 25 & 11:12:05 &  18:18:02  &  5 &  5700 & N  &  1 2 3\nl
 178& Jun 26 & 10:50:03 &  17:56:00  &  5 &  5220 & N  &  1 2 3\nl
 179& Jun 27 & 15:30:57 &  17:34:22  &  3 &  2770 & M  &  - 2 3\nl
 181& Jun 29 & 11:27:23 &  15:11:55  &  3 &  2490 & M  &  - - 3\nl
 220& Aug 07 & 11:25:19 &  18:37:51  &  5 &  5750 & N  &  1 2 3\nl
 221& Aug 08 & 10:05:49 &  17:17:19  &  5 &  4740 & N  &  1 2 3\nl
 222& Aug 09 & 10:49:03 &  17:52:31  &  5 &  4850 & N  &  1 2 3\nl
 223& Aug 10 & 12:13:19 &  19:16:15  &  5 &  5360 & N  &  1 2 3\nl

\enddata
\tablenotetext{}{DOY = day of year, M = medium mode, N = Nominal mode}
\end{deluxetable}
The most compelling evidence for the existence of a black hole in Galactic
X-ray binaries normally comes from the measured mass function which indicates
that the mass of the compact object is much
larger than that permitted for a neutron star. In the absence of measured
binary parameters (like in the case of GRS~1915+105) phenomenological 
arguments are normally used, which, though compelling for a class of 
objects, are not conclusive enough for individual cases. This is mainly due
to the fact that the accretion disk around a black hole has properties
quite similar to that around a low magnetic field neutron star (\cite{tana:95}).
Recent progress in the understanding of accretion onto black holes 
has suggested that  the black
hole accretion disks are cooled by advection in their innermost parts  
(\cite{chak:96a}; \cite{abra:97}; \cite{nara:98}).  Based on the new 
accretion theories involving advection, features
in black hole accretion which uniquely distinguishes them
from low magnetic field accreting neutron star,  have been 
identified. Narayan et al. (1997a) have  argued that advective 
cooling can occur throughout the disk for black hole accretion
providing  a unique way of identifying black hole
binaries in their quiescent state. Chakrabarti and Titarchuk (1995)
have argued that in the very high state of the sources, black hole
binaries should have a unique extended power law due to bulk comptonisation
(see also Laurent and Titarchuk 1998). 
In an earlier paper (\cite{paul:98b}, hereafter Paper I), we presented a possible evidence for the direct
detection of advection in GRS~1915+105. This is based on the detection of
regular and persistent X-ray bursts  which have a slow exponential rise,
 sharp decay and hardening of the spectrum as the burst progresses.

In this paper, we present a detailed analysis of all the IXAE observations
of GRS 1915+105 during 1997 June-August. We specially study  
temporal variations on a time scale from few seconds to few minutes. 
In the following sections, we 
describe the observations and the properties of 1889 bursts observed with PPCs.
We discuss our results  in the framework of advective  accretion disk
models. 

\section{Observations}

The Indian X-ray Astronomy Experiment (IXAE) onboard the Indian satellite 
IRS-P3 consists of three identical pointed proportional counters (PPCs) and one X-ray sky monitor and it was launched on 1996 March 21 from Shriharikota Range, India. The observations were carried out using all the 3 PPCs of IXAE. 
The PPCs are filled with argon-methane mixture at 800 torr pressure and 
have a total area of 1200 cm$^2$. The operating energy range is between 
2 keV and 18 keV and a passive collimator restricts the field of view to 
$2.3^\circ \times 2.3^\circ$. The energy resolution is 
$\approx 22{({E\over6})}^{-{1\over2}}\%$ at E keV with a detection 
efficiency of about 65\% at 6 keV and 10\% at 15 keV.  Each PPC is a 
multilayer unit consisting of 54 anode cells of size 1.1 cm $\times$ 1.1 
cm arranged in 3 identical layers.  The end cells  of  each layer and all 
the 18 anodes of the third layer are connected  together and  operated as 
a veto layer for the top two layers which  constitute the X-ray detection 
volume. The alternate anodes in each of the two X-ray detection layers are 
joined together and operated in mutual anti-coincidence to reject charged 
particle induced background.  Each PPC has its own front-end electronics 
and a processing electronics.  The processing electronics selects the 
genuine events based on the pre-determined logic conditions. An 8086 
microprocessor based system handles the data from each PPC and stores them in 4 Mbits 
of memory. The data storage is done in different modes which can be set 
by commands.  For further details of the PPCs and the observation 
methodology see Rao et al. (\cite{rao:98}).  

The IRS-P3 satellite is in a circular orbit  at an altitude of 830 km 
and inclination of $98^\circ$.  A star tracker onboard the 
IRS-P3 satellite co-aligned with the viewing axes of the proportional 
counters is used for pointing towards the X-ray sources with a pointing
accuracy of about $0.1^\circ$.  
The high inclination and high altitude orbit is found to be very background
prone and the useful observation time is limited to the latitude ranges 
typically from $-30^\circ$ S to $+50^\circ$ N. Further, the South Atlantic
Anomaly (SAA) region restricts the observation to about 5 of the 14 orbits 
per day. Observations with the PPCs are usually made in about 5 orbits of the 
satellite every day  in the nominal mode (N) with 1.0 s time resolution 
and each observation has a duration of about 20 minutes.  In the medium 
mode (M) with time resolution of 0.1 s, data are usually available only for 
three orbits due to the limited size of the onboard data storage unit. 
During 17 days of observations from June 12 to August 10, data from 71 
orbits were collected and  a total of 67,750 seconds of useful exposure 
time was obtained. A summary of the observations is given in 
Table 1. 

\placetable{tbl-1}

\section{The burst profiles}

Intense X-ray  bursts are observed through out the PPC observations over the
period of 1997 June 12-29 and August 7-10.  These bursts can broadly be 
put into two classes: regular bursts
lasting typically for a few days and centered around a fixed period 
with  low dispersion ($\delta P / P \sim 5 - 10 \%$)
and irregular bursts with no fixed periodicity ($\delta P / P \ge  50 \%$).
The  period of regular bursts shows two distinct time scales 
during our observations and quasi-regular bursts with properties 
in between those observed for the regular and irregular bursts have
also been seen. The observed bursts are, therefore, classified into four 
types:  (a) regular bursts having a 
slow rise and fast decay lasting for $\sim$ 15 s and recurring every 21 s, 
(b) regular bursts, having a slow rise and sharp decay lasting for 
$\sim$ 20 s and recurring every 46 s,
(c) quasi-regular bursts of variable duration, slow rise and sharp
decay, and (d) irregular  bursts, with duration of a few tens to a few hundred
seconds, followed by sharp decay. Sharp decay is a common feature of all
the bursts.  All the regular bursts  
usually have two peak structure while quasi-regular and irregular bursts 
show multi-peak structure.   

\placefigure{fig1}

Representative 2$-$18 keV light curves, added for all the 
PPCs except on June 27 when only PPC 2 \& 3 were on, of 300 s duration 
obtained on different days are shown in top four panels of Figure 1. 
These burst profiles were 
detected independently in each of the PPCs. All the panels in the figure have 
similar Y-axis scales.  Regular bursts of $\sim$ 21 s recurrence time 
were detected during August 7-10 with second peak being prominent and 
a sharp narrow dip between the two peaks; regular bursts of $\sim$ 46 s 
recurrence time were detected during June 12$-$17 and again during June 
22$-$26 with first peak being prominent; quasi-irregular bursts were seen 
during June 19$-$21 and irregular bursts were detected on June 18 and 
June 27-29.   In the  second panel (from bottom), we show regular bursts 
observed on 1997 June 22. The time zero corresponds to 1997 June 22, 19:10 
UT with the PPCs.  Similar burst profiles are seen from publicly available 2$-$13 keV 
RXTE/PCA data of 1997 June 22  which are shown in the bottom panel of 
Figure 1. The time zero corresponds to 1997 June 22, 19:35 UT. Remarkable
similarity between temporal profiles of regular bursts observed by PPCs and
independently by the PCA about 25 minutes later is apparent from the first and
second panels (from the bottom) of Figure 1.
A secondary peak near the end of the bursts is a common feature of all the 
bursts.   The quasi-regular and irregular long bursts show higher variability 
near the end of the burst and the burst duration is correlated to the 
quiescent state period just prior to the burst which we shall discuss 
in detail in the next section.

\begin{deluxetable}{lcccccl}
\footnotesize
\tablecaption{Summary of  characteristics of different
 type of bursts from GRS~1915+105. \label{tbl-2}}
\tablewidth{0pt}
\tablehead{
 \colhead{Type of}& \colhead{Mean recurrence} & \colhead{No of}   & \colhead{Norm.\tablenotemark{a}} & \colhead{Mean q.} &\colhead{Mean b.}& \colhead{Date of} \\
 \colhead{burst} & \colhead{time or interval (s)} & \colhead{bursts} &\colhead{q. flux} & \colhead{H R\tablenotemark{b}} & \colhead{H R\tablenotemark{b}} &\colhead{observation}
}
\startdata
reg. bursts &  21 $\pm$3 &995&2.87&1.28$\pm$0.05&2.2$\pm$0.2& Aug 7-10\nl
reg. bursts &  46 $\pm$5 &738&1.75&0.98$\pm$0.02&1.2$\pm$0.1& Jun 12-17 \&\nl
&&&&&& Jun 22-26\nl
quasi-reg. & from 50 to 90    &115&1.2 &0.94$\pm$0.02&1.5$\pm$0.1& Jun 19-21 \nl
bursts&&&&&&\nl
irreg. bursts & from 18 to 350   &67 &1.0 &0.94$\pm$0.02&1.5$\pm$0.1& Jun 18 \&\nl
&&&&&& June 27-29\nl
\enddata

\tablenotetext{}{q. = quiescent, b. = burst and H R = hardness ratio} 
\tablenotetext{a}{Normalised quiescent flux is normalised to irregular burst}
\tablenotetext{b}{HR is defined as the ratio of counts in 6$-$18 keV to that in 2$-$6 keV bands}
\end{deluxetable}

All the bursts start with  a well defined sharp peak and
decay faster than they rise. We define a burst event as a full cycle
of one quiescent interval followed by one burst. The recurrence time is
the sum of quiescent time and the burst duration. With this definition there is
no interval between burst events. This defines the individual burst events 
adequately for the regular and quasi-regular bursts. In the case of 
irregular bursts, we put additional criterion that  separate events are 
considered to be only those whose quiescent count rate goes below 250 counts 
s$^{-1}$ for individual PPC in the total 2.0-18.0 keV energy band. 
We measured the start of a 
burst event (corresponding to the end of the previous one) as the time of the
small dip at the end of the decay. Since all the bursts start with a sharp
peak, the time of the peak can be taken as the separation between the 
quiescent phase and the burst phase. 
We have marked the positions in the third panel of Figure 1 for a burst event: 
beginning of quiescent phase (preceding) is marked by 'a', end of the quiescent
phase and start of burst phase is marked by 'b', and end of the burst phase and
start of next quiescent phase (following) is marked by 'c'.  
We have detected bursts in all our 
observations. We have calculated mean recurrence 
time for each day and results are shown in Figure 2. The error
bar represents one $\sigma$ variation in the recurrence time during the
observations on each day. The large variations on June 18, June 27, and
June 29 represent irregular bursts  on these days while small variations 
on June 12-17, June 22-26 and August 7-10 show regular bursts during these
durations. The quasi-regular bursts were observed on June 19 and June 21.

\placefigure{fig2}

 In all the bursts, a dip is present just 
before the decay of the burst. But the most remarkable feature of our 
observations is the persistence of the regular bursts for a few days with 
almost similar shape, structure and period.  For both types of 
regular bursts, the recurrence time for 
the successive bursts shows a random walk in time 
instead of any regular pattern.  The distribution of burst recurrence time 
for each day fits well with a Gaussian, with a tail on the higher side, 
having a width $\sigma$ of  3-5  s for both  the types of regular bursts. 
 In the case of irregular bursts, the distribution of burst recurrence time 
shows  large variations.

\placefigure{fig3}

To improve the statistical accuracy of the data we have co-added
a large number of bursts by matching the last peak.  
The co-added burst profiles in two different energy
ranges (2$-$6 keV and 6$-$18 keV) are shown in the top  panels
while the hardness ratio is shown in the bottom panels of Figure 3 (a \& b) for 
all four types of bursts.
We chose to align individual burst of the same type to the last peak in order to keep sharp features during 
the decay of all the bursts while the sharp features during the rise are smeared
out due to the addition of bursts of different duration specially in the case of quasi-
regular and irregular bursts. Intensity changes
are more prominent at higher energy and the energy spectrum becomes
harder as the burst progresses in all types. The burst is hardest near the end
of its decay. This is a unique feature of these bursts which
distinguishes them from the bursts seen in LMXBs which become
softer in the decaying phase (\cite{lewi:95}). 

\placefigure{fig4}

We show the `rise' and the `decay' segments of  the profile of different type 
of the bursts in Figure 4.   We arbitrarily chose the burst start time at 0 s 
and the burst end time at 40 s for all types of the bursts to highlight the 
fact that  the slow rise and sharp decay is a common feature of all 
the bursts. The flux is normalized to the start point for the rise segment 
and to the end point for the decay segment of the burst profile.  
The `rise' and the `decay' profiles are strikingly similar for both the types of 
regular bursts observed on August 9 and June 22.    A least square fit with a function $f(t)= a \times exp^{((t-t_0)/t_r))}$ to the rising segment of the burst profile gives time constant $t_r$ = 10 s  for the regular bursts while it has a value in the range of 5 s to 6 s  for the quasi-regular and irregular bursts.
The burst decay is consistent with a straight line fit which gives a time
constant (defined as  the time required to drop from twice the quiescent flux)
in the range of 3 s to 7 s.

\section{Burst statistics}

A summary of the characteristics of different type of bursts  
is  given in Table 2.  Also given in the table are  mean quiescent and 
mean burst hardness ratio, defined as the ratio of counts in 6$-$18 keV to 
that in 2$-$6 keV.
A total of 995 regular bursts of $\sim$ 21 s 
recurrence time  (in $\sim$ 20,700 s of observation), 738 regular bursts 
of $\sim$ 46 s recurrence time (in $\sim$ 33,560 s of observation), 115 
quasi-irregular bursts (in 6,600 s) and 67 irregular bursts (in $\sim$ 
6,890 s) have been detected.  The  peak intensity varies from 1.5 to 3.5 
times the quiescent intensity.

\placetable{tbl-2}
\placefigure{fig5}

In the case of quasi-regular and irregular bursts, the burst duration is 
correlated to the quiescent time just prior to the burst.   
We have measured the quiescent time and the burst duration for all types of 
bursts according to the definition given in the previous section. Results 
are shown in Figure 5 
along with a least square straight line fit  which shows a good correlation 
between the burst duration and the preceding quiescent time for the 
quasi-regular and  irregular bursts. 

\placefigure{fig6}

We do not see any such correlation for the regular bursts.   Inclusion 
of the regular bursts reduces the correlation coefficient  from 0.94 to 0.83 
(shown in the inset of Figure 5).  Similar plots for the regular bursts of 
mean recurrence time  21 s  and 46 s are shown in the top and the bottom 
panels of Figure 6 respectively for three days in each case. 
The dotted line is a least square fit to the quasi-regular and irregular 
bursts and is shown here for comparison. For both the types 
of regular bursts, the burst duration is constant for each day but it 
does show day to day variation.   It may be stressed here that although 
the regular bursts themselves do not show  any correlation between the 
preceding quiescent time and the burst duration, they fall very close to 
the relation derived for the quasi-regular and irregular bursts (see 
inset of Figure 5).  We do not find any correlation between the quiescent 
time following the burst and the burst duration and results for 257 
different type of bursts are shown in Figure 7.  The observed correlation 
for the burst duration and the preceding
quiescent time for the irregular and quasi-regular bursts
could be simply a reflection of the fact that the  
burst duration and the quiescent time are of comparable magnitude for 
such bursts. 
This correlation, however, establishes that a given burst cycle
starts at the beginning of a quiescent phase and gets completed at 
the end of a burst phase because we do not find any correlation
between the burst duration and the following quiescent time. 
Henceforth, we shall use 
the preceding quiescent time as the quiescent time of a burst.   It is 
interesting to note  that we have observed continuously the full cycle 
of regular - irregular - quasi-regular and again regular bursts from 
June 17 to June 22, 1997. 

\placefigure{fig7}
\placefigure{fig8}

We have calculated the average flux during the burst phase and during the quiescent 
phase separately for all the different type of observed bursts and results 
are shown in Figure 8. The ratio of average flux during the burst time to 
the average flux during the quiescent phase is plotted as a function of 
the burst duration. For quasi-regular and irregular bursts, this 
ratio shows good correlation with the burst duration (least square fit 
straight line) and has a value of two and higher. This ratio is, however, 
constant and has a value of less than  two for the regular bursts as shown 
in  the inset of Figure 8.   The dotted line shows the least square 
straight line fit for the quasi-regular and irregular bursts. 

\placefigure{fig9}

We have calculated average hardness 
ratio during the quiescent phase for all types of bursts and results 
are shown in Figure 9. The hardness ratio decreases as the quiescent time 
increases and it is higher for the regular bursts than that for the 
quasi-regular and irregular bursts.  The ratio of average flux during 
the burst time to the average flux during the quiescent phase is plotted 
in Figure 10 as a function of the  average hardness ratio during the 
quiescent phase for  all the observed bursts. The solid line is a 
least square fit to both 
types of regular burst data (442 regular bursts) which shows good correlation. 

\placefigure{fig10}

\section{Discussion}

Because of their unique feature of slow rise and fast decay, the bursts 
in GRS~1915+105 are very different from the type I
X-ray bursts seen in about 40 LMXBs and type II X-ray bursts 
in the Rapid Burster (MXB 1730$-$335). All the bursts in the LMXBs have
fast rise time of less than a second to a few seconds and slow
decay of 10 seconds to a few minutes (\cite{lewi:95}).
The type I X-ray bursts are understood to be
thermo-nuclear flashes caused by accretion of matter on to the surface of
the neutron star. The type II bursts are produced by sudden infall of
matter on to the neutron star due to some instability in the
inner part of the accretion disk supported by the magnetic field.
The slow decay of the burst intensity represents the cooling time 
scale of the neutron star photosphere.   In the classical
bursts, the spectrum is initially hard and becomes softer as
the burst decays (\cite{lewi:95}). In sharp contrast, the
bursts in GRS~1915+105 remain hard till the end
and it is, in fact, the hardest near the end of the burst. 
 
 In the case of type I X-ray bursts, the ratio of the burst 
luminosity ($L_b$) and
the average quiescent X-ray luminosity ($L_p$) is
$ {L_b\over L_p} \sim 10^{-2}$. On the other hand, the time-averaged
type II burst luminosity is much higher, usually 0.4 to 2.2 times the
average luminosity of quiescent emission (\cite{lewi:95}).
The time-averaged luminosity of the regular bursts detected from
GRS~1915+105 is from 0.15 to 0.9 times the luminosity of the quiescent
emission. This is much higher than the ratio in type I bursts
(where the thermonuclear process has much smaller efficiency compared
to the gravitational process) and less than the type II bursts (where
the burst emission is due to gravitational energy release).
The emission process involved in producing the bursts here is not
likely to be thermo-nuclear because of the energetics involved.
If the energy generation process is gravitational (like in type II
bursts), the difference in efficiency might indicate the absence of hard 
surface in the compact object.  A process in which the energy produced 
is due to gravitational potential but not all the energy is emitted as 
radiation, part of it being advected into the event horizon as kinetic 
energy of the matter, is appropriate for this source. This probably 
provides a compelling evidence that the compact  object in GRS~1915+105 
may be a black hole.

The quasi-regular and irregular bursts show higher variability near the 
end of the burst as we noted earlier and the burst duration is correlated to the quiescent time.  
Similar behavior is also reported from the PCA observations carried out in 1996 
June (\cite{bell:97a}). Several irregular bursts, concurrent with the 
present observations on 1997 June 18 and having similar properties, have
also been detected in the PCA data (\cite{bell:97b}). They modeled these 
bursts as a consequence of emptying and replenishing of the inner accretion
disk caused by a viscous thermal instability. 
Paul et al. (1998b), on the other hand, attempted to explain the
regular bursts as due to periodic in-fall of matter onto a black hole
from an oscillating shock front. In the following we attempt to give
a comprehensive picture of the various types of bursts
observed in GRS 1915+105 in the light of the recent theories
of advective accretion disks.

\subsection{Advective accretion flows around black holes}

Recent work on the theory of accretion onto black-holes
(see \cite{abra:97}; \cite{chak:96b}; \cite{nara:98}
 for reviews) has shown that
advection cooling is
important in the innermost part of the accretion disk. For 
hot optically thin disks strong advective cooling occurs everywhere, very 
far from the black hole as well.
\cite{nara:94}  have taken a self-similar solution and 
have divided the possible solutions into two branches: the first
type where the energy is trapped from the disk and converted
to jets and the second type with advection dominated thick
accretion disk. A few observed sources are compared with the
predictions of the Advective Dominated Accretion Flows (ADAF)
 with a fair degree of success (\cite{nara:97b}).  \cite{chak:95}, 
 on the other hand, have taken a complete solution of 
viscous transonic equations and demonstrated that the accretion disk 
has a highly viscous Keplerian part which resides on the equatorial 
plane and a sub-Keplerian component which resides above and below it. 
The sub-Keplerian component can form a standing shock wave (or, more 
generally, a centrifugal barrier supported dense region) which heats 
up the disk to a high temperature.  The need to define the viscosity 
parameter is circumvented by taking two accretion rates: the accretion 
through the classical ``standard'' disk and the accretion through the 
sub-Keplerian component. \cite{ebis:96}  have attempted to explain 
the observed X-ray spectrum, particularly the change of spectral states 
in the black hole candidates, using this model.
In both the models the change of spectral states are  ascribed to the
change over from a purely thin accretion disk (with advection 
occurring very close to the black hole) in the high-soft state to 
the advective disk extending over a large distance in the low-hard state.
The hard X-ray power-law component is ascribed to the 
Comptonisation spectrum from the advective disk and the 
Shakura-Sunyaev multi-temperature disk emission (which
is predominant in energies below $\sim$ 10 keV)
is associated with the standard thin disk.
In the ADAF model of \cite{nara:94} the advective thick disk changes 
into a standard thin disk at a distance r$_{tr}$ whereas according
to \cite{chak:95}  the advective thick disk and the
standard thin disk co-exist upto a certain radial distance and
a standing shock wave or a centrifugal barrier dominated dense region is
a common feature of the sub-Keplerian component.
In the following we try to examine the burst properties of GRS 1915+105
in the light of these new accretion disk theories incorporating advection.

The source was in a low-hard state during 1996 December to 1997 March 
(\cite{grei:98}) when the hard X-ray spectral index ($\sim$2.0) and 
the soft X-ray flux (300 - 500 mCrab) were low. The source started a 
new outburst around 1997 April-May when the soft X-ray flux started 
increasing and the X-ray spectrum  became soft (spectral index increased 
to 3 $-$ 4).  We suggest that the peculiar bursts that we have seen 
are characteristics of the change of state of the source.

In the quiescent state of the burst the source is in the hard state. 
This is evident from the large value derived for the 
inner disk radius (R$_{in}$) of the multi-temperature thin-disk model
fitted for the energy spectrum. The fitted value of R$_{in}$ is around 
300 km for very long bursts of 1000 s duration
 (\cite{bell:97a}) and 30 -  100 km  for
the irregular  bursts (\cite{bell:97b}) 
and 30 km for regular bursts (\cite{taam:97}).   Note that the
derived values for the radius can be underestimate due to scattering 
effects and also due to the approximation involved in fitting 
the Comptonised part of the spectrum as a power-law (see Shrader \& 
Titarchuk 1998).  It should be further noted here that the derivation 
of R$_{in}$ is
very much model dependent and we use this quantity only for a 
qualitative description of the spectral states and also to make 
an order of magnitude estimate of time scales involved  in the change 
of spectral states.
The fitted temperature of the disk is 1 - 1.5 keV.  
The spectral index of the power-law component during the quiescent state 
of the long bursts is $\sim$ 2.22 (\cite{bell:97a}) indicating that the 
source is truly in a hard state.  During the burst phase the intensity 
is higher, the radius of the disk is smaller (20 - 30 km), the temperature 
is higher (2 - 3 keV) and the power-law index is steeper (3.57 for the 
long bursts and 3.3 for the regular bursts). These characteristics 
strongly suggest that the source is in a high-soft state during the 
bursts.  Hence there are strong indications  that the source 
makes state transitions in very short time scales corresponding to the 
rise and fall time of the bursts (a few seconds). Such fast changes of 
states are possible in the two component accretion flows where the 
advective disk covers the standard thin disk (Chakrabarti 1996b).   In 
the following we describe the bursting behaviors of GRS 1915+105 within 
this scenario, taking the model parameters given in  \cite{chak:95}  and 
also in Narayan et al. (1998).

 In the low-hard state of the source the thin Keplerian disk is visible
only from a large  radial distance R$_o$, the sub-Keplerian component completely
encompasses the thin disk below this radius (the soft photons from
the disk act as  seeds for the Comptonisation process).
 When the disk accretion 
rate ($\dot{m_d}$) increases, at some critical point, the non-Keplerian
halo accretion rate ($\dot{m_h}$) can decrease and the high-soft state can set in. 
This change in accretion rate can occur either due to the change in
the total accretion rate ($\dot{m_t}$ = $\dot{m_d}$ + $\dot{m_h}$) or due to 
some changes in the viscosity in the thin accretion disk which
changes  $\dot{m_d}$ (keeping $\dot{m_t}$ constant).
We suggest that when the total accretion rate is close to some critical value
the source can change states and the observed irregular bursts are the
manifestations of such changes of state.

To understand the mechanism of the bursts let us equate R$_{in}$ to R$_o$ 
from where the advection
dominated halo component starts covering the thin accretion disk in 
the low-hard state of the source (the quiescent state of the burst). 
Assume that $\dot{m}_d$ is close to a critical value where the
change of the state takes place. At some particular point of time the
boundary condition at the inner edge changes such that $\dot{m}_d$
increases. The spectral state, however, will remain unchanged till
this effect reaches R$_o$. The time scale for this to happen is the
viscous time scale of the thin accretion disk.  Assuming the standard 
$\alpha$ disc we can write the viscous time
scale as R$_o^2 / \nu$ where $\nu$ is the viscosity coefficient which
is given as $\alpha c_s H$ for the $\alpha$ disk, where $c_s$ is the
sound speed and H is the disk thickness, and $\alpha$ is the
viscosity parameter. Taking the scaling laws
for H and $c_s$ (\cite{fran:85}), we can write the viscous
time scale of the disk as 
\begin{equation}
t_{vis}^d = 4.3 \times 10^{-4} \alpha^{-1} \dot{m}^{-1}_d m^{-1} R^2_o ~~ s
\end{equation}
where 
$\dot{m}_d$ is in the units of Eddington accretion rate and m is the mass of the
black hole in solar mass units and R$_o$ is in km. Substituting 
$\dot{m_d}$ = 1, m = 10, $\alpha$ = 0.01, and R$_o$ = 300, 
we get t$_{vis}$ = 400 s,
agreeing with the observed quiescent state time scale of long bursts.
This also explains the non-linear dependence of burst quiescent time
with radius reported by Belloni et al. (1997b) for the long irregular bursts.

At R$_o$, the increased $\dot{m}_d$ decreases  $\dot{m}_h$ and the
advection dominated halo component completely advects onto the black 
hole in the viscous time scale of the halo component given by 
\begin{equation}
t_{vis}^h = {{R_o^2}\over{H \alpha c_s}}  
\end{equation}
  Here we assume that the halo component is an advection dominated
accretion disk in which  the temperature can go very high (see Narayan et al.
1998). For 
advection dominated thick disks we can take H $\sim$ R and use
the scaling law for c$_s$ as
\begin{equation}
c_s = 1.18 \times 10^{10} (R_o/R_s)^{-{1\over{2}}} cm s^{-1} 
\end{equation}
where R$_s$ is the Schwarzchild radius (\cite{nara:98}).
We can rewrite the viscous time scale as
\begin{equation}
t_{vis}^h = 4.9 \times 10^{-6} \alpha^{-1}  m^{-{1\over{2}}} R^{3\over{2}}_o ~~ s
\end{equation}
Substituting the values as earlier, we get 
$t_{vis}^h $ as  $\sim$ 1 s, which is quite close to the observed
rise time of the burst.
Now the burst phase starts, which is nothing but the soft state of the
source with R$_o$ coming very close to the shock front (or the
centrifugal barrier supported dense region). Note that the value of 
R$_o$ derived by spectral fitting is always 20 - 30 km during the
burst maximum, irrespective of
the type of the burst. At some particular time the inner boundary condition 
can change again and $\dot{m}_d$ can decrease and 
$\dot{m}_h$ can increase and this sub-Keplerian component
can suppress inner part of the accretion disk and a Compton cloud can
be generated. In the advection dominated accretion flow the radial velocity,
typically, will be
\begin{equation}
 v = -1.9 \times 10^{10} \alpha R_o^{-{1\over{2}}} m^{1\over{2}} ~~~~ cm ~~ s^{-1} 
\end{equation}
(see \cite{nara:98}). For a radius of 300 km, this will have a 
time scale of $\sim$ 1 s, which is seen as the fast decay time of the bursts.

For smaller values of R$_o$, the in-fall time scale will be lower 
(it goes as R$_o^{{3\over{2}}}$) and for some particular value of 
R$_o$ this time scale can match the cooling time scale of the post
shock region, which is about 0.01 s for  a  shock radius 20 r$_g$ 
(where r$_g$ is the Schwarzchild radius) and black hole mass of 10 
M$_\odot$ (\cite{paul:98b}).  When the two time scales match they 
can give rise to oscillations which are quasi-periodic in nature 
(\cite{molt:96}).  The in-falling
matter can immediately trigger the instability at the inner boundary
of the disk and the source can immediately revert back to the hard state.
These regular oscillations
accumulate matter at the shock-front (or the centrifugal barrier)
and they can catastrophically fall onto the black-hole.  The scenario 
described above explains in a qualitative way
the various types of bursts observed in GRS 1915+105. 
We must mention here that the time scales and sizes that we have taken 
for the calculations are very approximate due to the uncertainties in the
multi-temperature disk parameters and the viscosity parameter.

The scenario sketched above is similar to that described by Belloni et al. 
(1997a) as
far as the viscous time scale is concerned. These authors, however,
make the assumption that when the source reverts back to the low intensity
state of the burst, the matter between R$_o$ and R$_i$ disappears
behind the black hole at a free fall time scale. It is worth pointing
out that a standard thin accretion disk cannot disappear at free-fall time scale
without effectively transferring its angular momentum and the
time scale for angular momentum transfer is the viscous time scale.
Further, the systematic change in the hard power-law index during the
bursts cannot be explained by only invoking a change in the
disk inner radius.

Paul et al. (1998b) have explained the regular bursts in terms of the 
oscillations of the shock front. For this to happen in the observed
time scale, the shock has to be very far away from the black hole. This
can only happen in the low state of the source. Actual observations,
however, suggest that the regular bursts occur when the source approaches
the soft-high state when the shock front is very close to the
compact object. Hence it appears that the periodicity of the
regular bursts occurs from the matching of the viscous time scales
rather than the shock front oscillation. The catastrophic in-fall
of matter at the shock-front, invoked by Paul et al. (1998b), seems to be
appropriate for the regular bursts.

Several of the correlations obtained in the preceding sections can be 
explained using the physical picture described above. 
The quiescent flux is related to R$_o$ (see Table 2). The rise time of 
the bursts represents the viscous time scales of the halo component.
The burst duration
is related to the preceding quiescent time (Figure 5) because 
the time scale for the inner boundary condition to change will
also be the viscous time scale, which is related to the R$_o$
of the preceding quiescent time. This correlation does not hold
for the regular bursts because the burst duration is 
independent of R$_o$ due to the resonance. During quasi-regular and 
irregular bursts, change in $\dot{m_d}$ is mainly due to change in
viscosity caused by a viscous - thermal - instability (see Figure 9; 
the variation in hardness ratio for these bursts during quiescent time
is much less than that in case of regular bursts) (\cite{bell:97a}). 
The high quiescent 
time X-ray flux during the regular bursts heats up the disk and 
suppresses the instability (\cite{lewi:95}).

The ratio of the average
fluxes during the burst and the quiescent time essentially
represents the ratio of the values for R$_o$ which is related to
the burst duration through the relation for the viscous time scale
(Figure 8). The hardness ratio (which represents the inner disk temperature)
shows weak dependence on the quiescent time for the irregular bursts because for
large enough R$_o$ the temperature does not change drastically with 
R$_o$ (Figure 9). The regular bursts of long duration (45 s) also have a
value for the hardness ratio during quiescent time which is similar
to that seen for the irregular bursts (though slightly higher). 
The variation in hardness ratio is larger which reflects the change in 
R$_o$ (see Figure 10). The mean burst hardness ratio for these bursts 
is $\sim$ 1.2 which suggests that these bursts occur during high 
state (\cite{nara:98}).  During the short duration regular bursts the 
source is almost in the soft-high state during the quiescent phase, 
showing a large value for the hardness ratio.   The mean burst hardness 
ratio is $\sim$ 2.2 which suggests very high state.  These observations 
imply that $\dot{m_d}$ is substantially larger in very high state than 
that in high state.  The flux ratio for the regular bursts represents 
the amount of amplification that is possible for the resonating bursts 
and it is related to the quiescent time temperature (Figure 10).

According to the scenario sketched above, the X-ray bursts are due
to fast changes in the spectral states when the source reaches the high 
state and  the accretion rate is very close to the Eddington accretion 
rate. Chakrabarti \& Titarchuk (1995) have pointed out that during such 
high state the source shows  distinct hard spectral component due to bulk 
motion comptonisation.  Shrader \& Titarchuk have fitted the high state 
spectrum of GRS 1915+105 using the bulk motion Comptonisation model and 
derived a temperature of 0.9 keV. Titarchuk \& Zannias (1998) have 
analyzed the exact general relativistic integrodifferential equation of 
radiative transfer for a realistic situation of accretion onto black holes 
in the high state and showed that an extended power-law spectrum results 
even from an arbitrary spectrum of low energy photons. Laurent \& 
Titarchuk (1998) have calculated the specific features of X-ray spectra 
using Monte-Carlo simulations and demonstrated the stability of power law 
spectral index over a wide range of mass accretion rate. This conclusion 
agrees with our observation that the hardness ratio during the quiescent 
time remains constant over  wide burst time scales (Figure 9), except for 
the regular bursts of 21 s duration which may be due to reduced optical 
depth (thus the halo) for these bursts. 

It is interesting to speculate the possible reasons
for the differences in the two regular bursts of $\sim$ 21 s
and  $\sim$ 46 s duration. As pointed out by Shrader \& Titarchuk (1998),
bulk motion Comptonisation sets in during the
soft-high state and the energy radiated is a very small fraction
of the accretion energy. Part of the energy can be used to drive
the matter away from the central source in terms of jets. The
difference in the two regular bursts may be due to the fact that
in the regular bursts of 21 s duration jet formation may be
setting in. It may be noted here that the time profiles of the
two types of regular bursts agree with each other but for a
sharp dip in the short regular bursts.

The ADAF models have been used to obtain unique identifying features 
of black hole sources (\cite{nara:97a}; Laurent \& Titarchuk 1998).    
An advection-dominated
accretion flow is one in which most of the energy released by viscous
dissipation is stored in the gas and advected to the compact object and a small
fraction of the energy is radiated.
The argument is that 
if accretion is via an ADAF and if the object has an event horizon, then 
the advection energy will disappear from sight. However, if the central 
object has a surface, then the surface will be heated by the hot inflow 
from the ADAF and the advected energy will be emitted as thermal radiation. 
This additional evidence for the black hole nature of GRS~1915+105 is 
qualitatively 
different from usual method that rely on a measurement of the mass. The 
usual argument is that if an object is too massive  to be a neutron star 
it must be a black hole.

\subsection{A test of the model using RXTE/PCA data}

In principle, one can test this model using X-ray data above 10 keV.  The 
Shakura-Sunyaev multi-temperature disk emission  dominates in 2$-$10 keV 
energy range while the power-law component dominates at higher energies 
(\cite{shra:98}).  We have analysed the publicly available RXTE/PCA data for 
irregular as well as regular bursts concurrent with our observations. 
The results of irregular bursts from RXTE/PCA  data of 1997 June 18 are 
shown  in Figure 11 along with the results from RXTE/PCA data of 1997 May 8 
and August 19 when GRS~1915+105 was in low state and in high state 
respectively.  Both the top panels and the bottom panels have same Y-axis 
scales. Similarly, both the left panels and the right panels (top and bottom) 
have same X-axis scales. This allows straight forward comparison of burst 
time data with the data during low and high state. GRS~1915+105 was in 
high state on 1997 August 19 and 2$-$13 keV flux varied from  19000 c/s 
to 30000 c/s while the source was in low state on 1997 May 8 and 2$-$13 
keV flux was steady around 6000 c/s (right top panel). No burst is observed 
during these observations. The ratio of 13$-$60 keV flux to 2$-$13 keV 
flux has a value of 12$\pm$2 \% during the low state on 1997 May 8 while it 
is $\sim$ 3\% during the high state on 1997 August 19 (bottom right panel).  
This implies  that the source  remains in the same state during these 
observations.

\placefigure{fig11}

The results of irregular bursts on 1997 June 18 are shown in left panels of 
Figure 11. The 2$-$13 keV flux varies from 4000 c/s to 28000 c/s which is 
shown in left top panel. These bursts have similar profile as those shown 
in second panel (from top) of Figure 1 from our data of 2$-$18 keV for June 27. The ratio 
of 13$-$60 keV flux to 2$-$13 keV flux is shown in left bottom panel. The 
relative flux of 13$-$60 keV reaches upto 12\% during quiescent time while 
it is  as low as  only 3\% during burst phase of irregular bursts. The 
minimum value of this ratio during burst phase is in  agreement 
with the value of this ratio for high state on 1997 August 19. Similarly, 
its maximum value during quiescent phase is in  agreement with the 
value for low state on 1997 May 8.   The 13$-$60/2$-$13 ratio drops to its 
minimum value of $\sim$ 3\% at the first peak of each burst which implies 
the cooling  of the halo component by the soft photons from the disk. It 
starts increasing immediately 
during the rest of the burst phase. This ratio reaches a value of $\sim$ 6\% 
during the burst decay phase which would mean a significant recovery of the 
halo component. The burst decay phase ,therefore, may represent unsaturated 
Comptonisation before the source reaches saturated level of low  state.

\placefigure{fig12}

Similar results of regular bursts from RXTE/PCA  data of 1997 June 22 are 
shown in Figure 12. The 2$-$13 keV 
flux varies from 7000 c/s to 29000 c/s (top panel).  The ratio of 13$-$60 
keV flux to 2$-$13 keV flux reaches upto 9\% during quiescent time and it 
drops upto  $\sim$ 3.5 \% during burst phase of regular bursts (bottom panel).  One would expect higher contribution from power-low component due to larger 
R$_o$ in case of irregular bursts than that for regular bursts. It would 
mean that resonance terminates the recovery of the halo component prematurely 
in the case of regular bursts. In fact, it is the major difference between the 
regular bursts and the irregular bursts as much longer time is available for 
the recovery of the halo component in the case of irregular bursts.  During 
bursts phase, the relative contribution of power-law component is lower in 
case of regular bursts as thin accretion disk extend to smaller radius 
and hence has higher temperature than that for irregular bursts 
(see Figure 10). 
The results of the other type of regular bursts of $\sim$ 21 s duration 
are also 
consistent with this scenario. The ratio of 13$-$60
keV flux to 2$-$13 keV flux reaches upto 6\% during quiescent time. The 
2$-$13 keV maximum flux during burst phase is same (approaching  a value of 
28000 c/s for 5 CPUs of RXRE/PCA) for all the types of 
observed bursts, provides another independent support to our model while
the quiescent time minimum flux varies almost by a factor of four.

\section{Conclusion}

The observed bursts from GRS~1915+105 are very different 
compared to the classical bursts in the LMXBs both in terms of temporal
structure and spectral evolution. 
Our results broadly put all the observed bursts in two classes: irregular and 
quasi-regular bursts in one class, and  regular bursts in another class. 
There is strong correlation 
between the preceding quiescent time and the burst duration for the 
quasi-regular and irregular bursts. No such correlation is found for 
the regular bursts. 
The ratio of average flux during the burst time to 
the average flux during the quiescent phase is high and variable in  
former case while it is low and constant in  latter case. 
We present a  comprehensive picture of the various types of bursts
observed in GRS 1915+105 in the light of the recent theories
of advective accretion disks.  We present a unified model for the origin
of these bursts which explains almost all the observed properties of these bursts. We suggest that the peculiar bursts that 
we have seen are characteristic of the change of state of the source. 
 The change of state is due to change in the disk 
accretion rate which may be either due to a change in the total 
accretion rate or due to some changes in the viscosity in the thin
accretion disk.  The periodicity of the regular bursts occurs from 
 matching of the viscous time scale with the cooling time scale of 
the post shock region.   We have presented a test of this model using 
13$-$60 keV RXTE/PCA data for irregular bursts and  regular bursts, during 
the low-hard state, and  during the high-soft state  which show good 
agreement with our model.   These results may be viewed as additional 
evidence that the X-ray source GRS~1915+105 is a black hole. 


We thank K. P. Singh for the valuable comments on the manuscript.   
We acknowledge K. Thyagrajan, Project Director of IRS-P3, R. N. Tyagi, 
Manager PMO  and staff of ISTRAC for their support during 
observations. The valuable contributions of the technical and 
engineering staff of ISAC and TIFR in making the IXAE payload are 
gratefully acknowledged. We thank RXTE team for making their data 
publicly available. We are grateful to the anonymous referee for
his constructive comments and suggesting improvements.


\centerline{\bf FIGURE CAPTIONS}
\figcaption[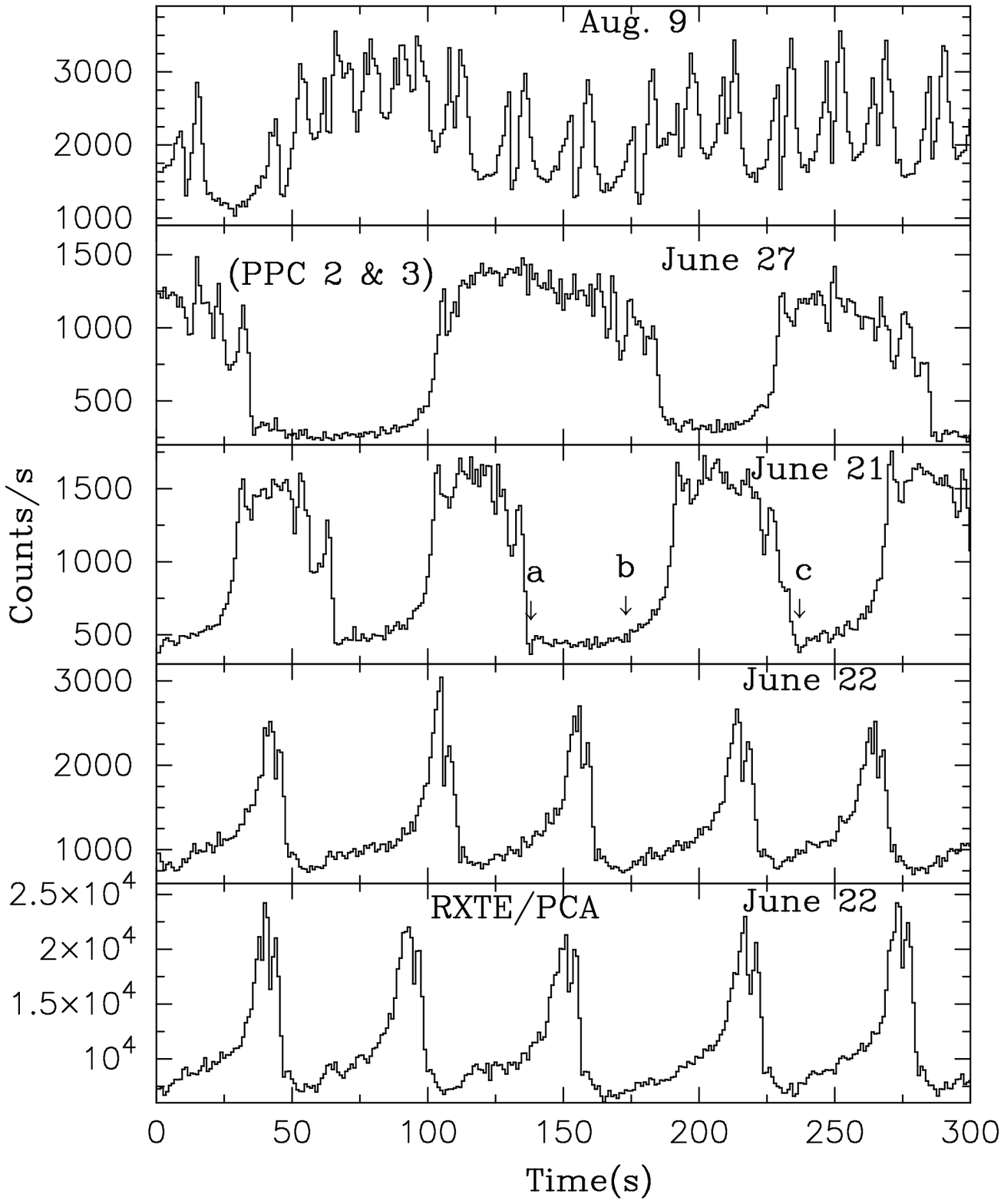]{The  regular bursts with $\sim$ 21 s recurrence time 
(first panel from the top),  irregular bursts (second panel), quasi-regular 
bursts (third panel) and regular  bursts with $\sim$ 46 s recurrence time 
(fourth panel)  observed in GRS~1915+105 with all the PPCs except on 
June 27 (irregular bursts) when only PPCs 2 \& 3 were on. Date of each 
observation is given in the respective panels. The regular bursts observed 
by RXTE/PCA on June 22 are shown in bottom panel for comparison with the 
regular bursts observed on June 22 with the PPCs.  For other details see 
in text. \label{fig1}}

\figcaption[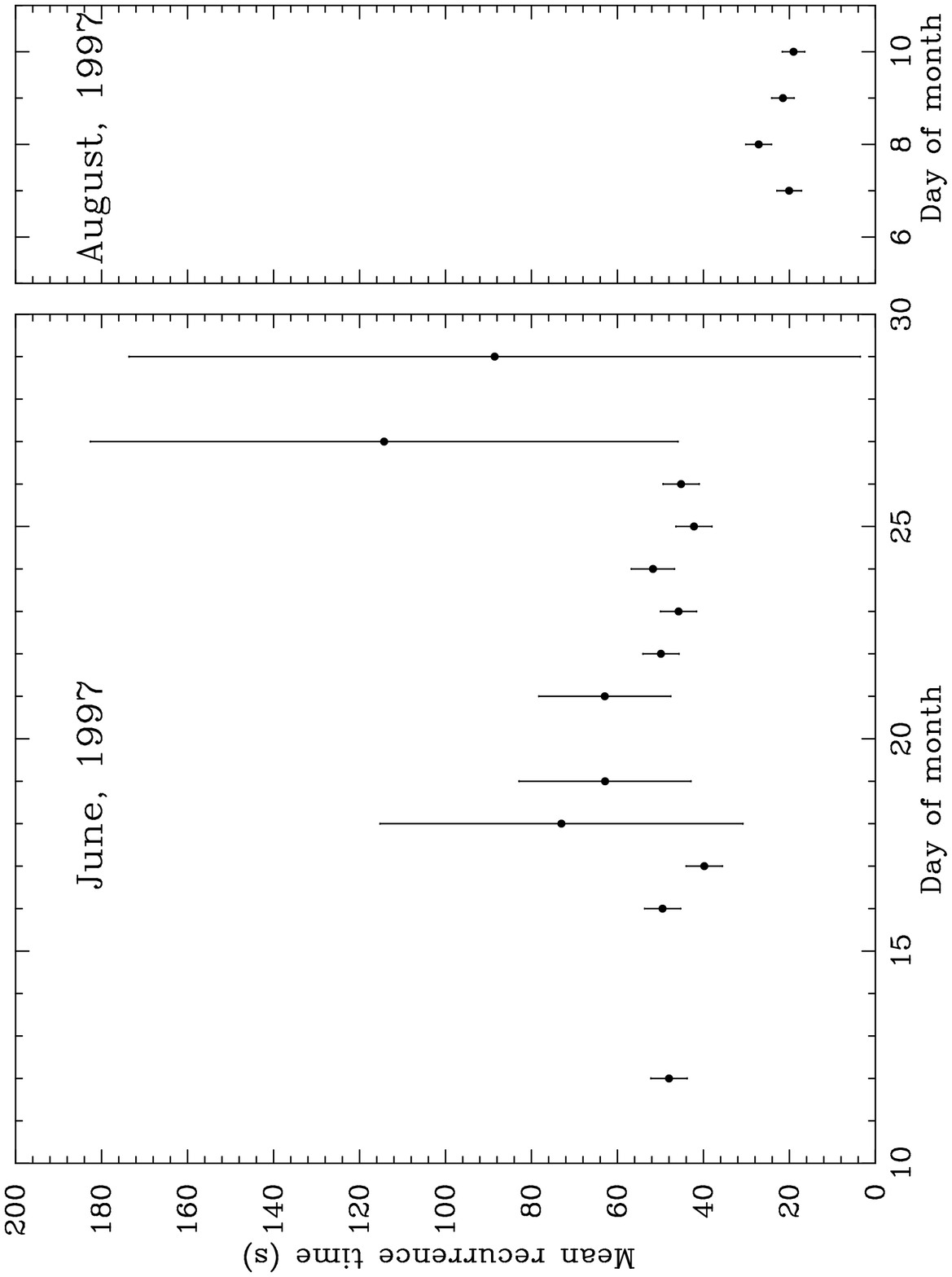]{The mean burst recurrence time for each day of observation. The error bars show 1 $\sigma$ variation in each day of observation. \label{fig2}}

\figcaption[fig3a&b.eps]{(a) The burst profiles in two different energy bands
are shown in the top  panels for the two types of regular bursts. The hardness ratio is shown in the bottom  panels and (b)     the burst profiles in two different energy bands are shown in the top  panels for the quasi-regular and irregular bursts. The hardness ratio is shown in the bottom  panels. \label{fig3}} 

\figcaption[fig4.eps]{The rise  and decay  segments of the burst profile
of all four types of observed bursts. The  start time of the bursts is
arbitrarily chosen at 0 s and the end time at 40 s for  clarity of comparison. The intensity of the rising segment is normalized to the start point while intensity of the decay segment is normalized   
to the end point. \label{fig4}}

\figcaption[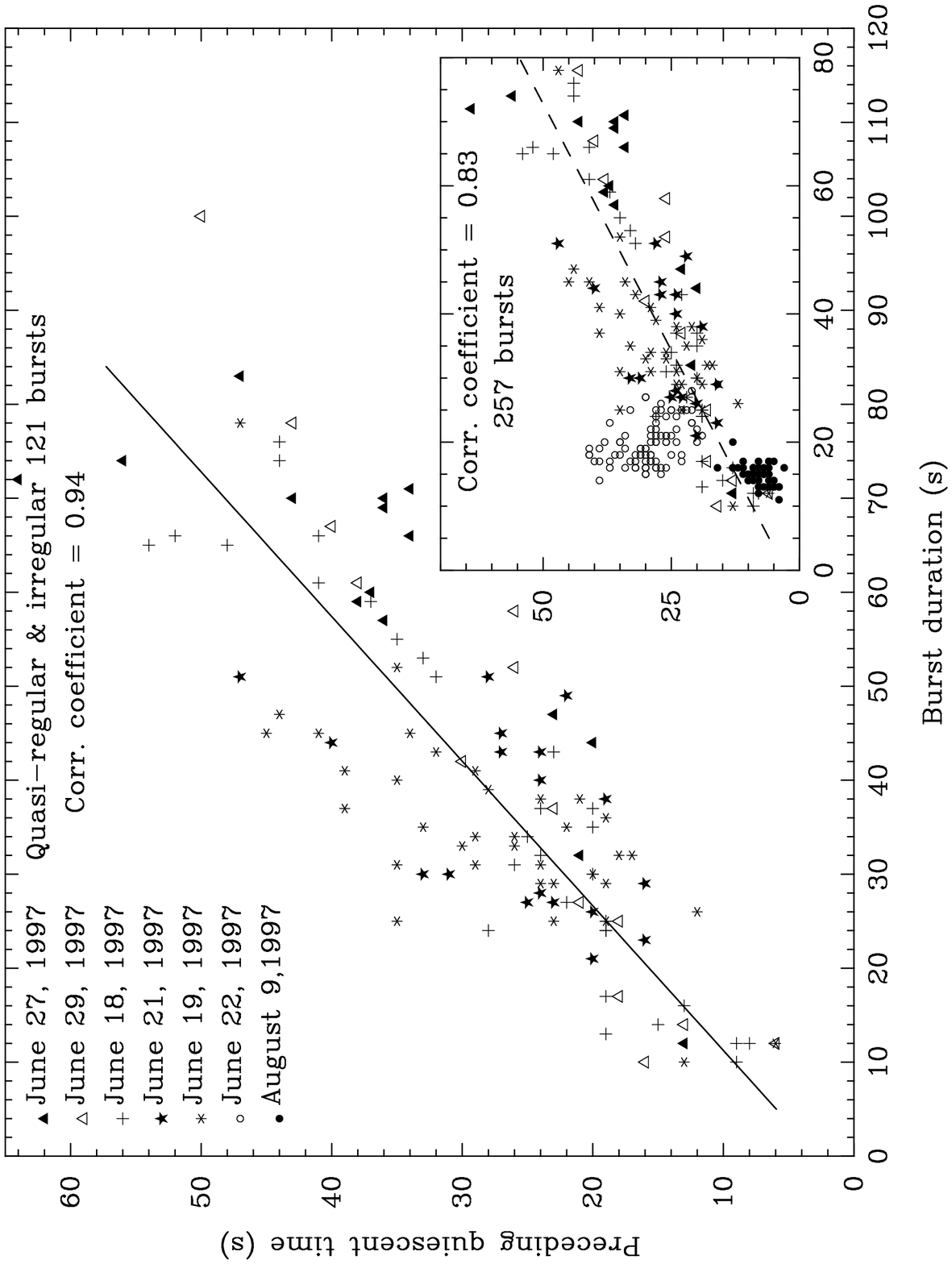]{Correlation between the preceding quiescent time and the burst duration (as defined in the text) for the quasi-regular and irregular bursts. The straight line is the least square fit to the data. In the inset, both types of regular bursts are also shown. \label{fig5}}

\figcaption[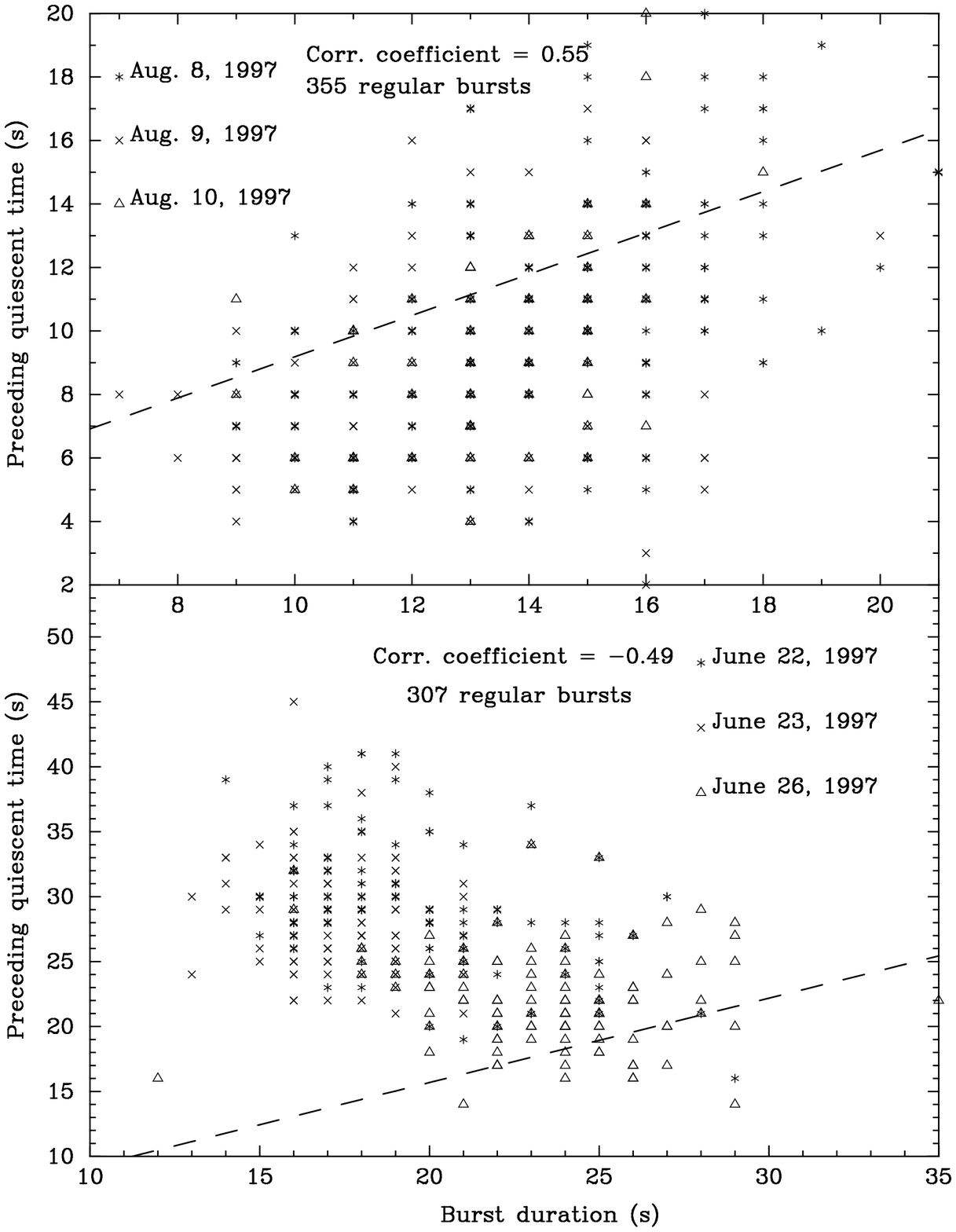]{Correlation between the preceding quiescent time and the burst duration for the regular bursts with $\sim$ 21 s recurrence time in the top panel and for the regular bursts with $\sim$ 46 s recurrence time in the bottom panel. Data are for three days in each case. The least square fit for the quasi-regular and irregular bursts (of Figure 5) is shown by dotted line for comparison. \label{fig6}}

\figcaption[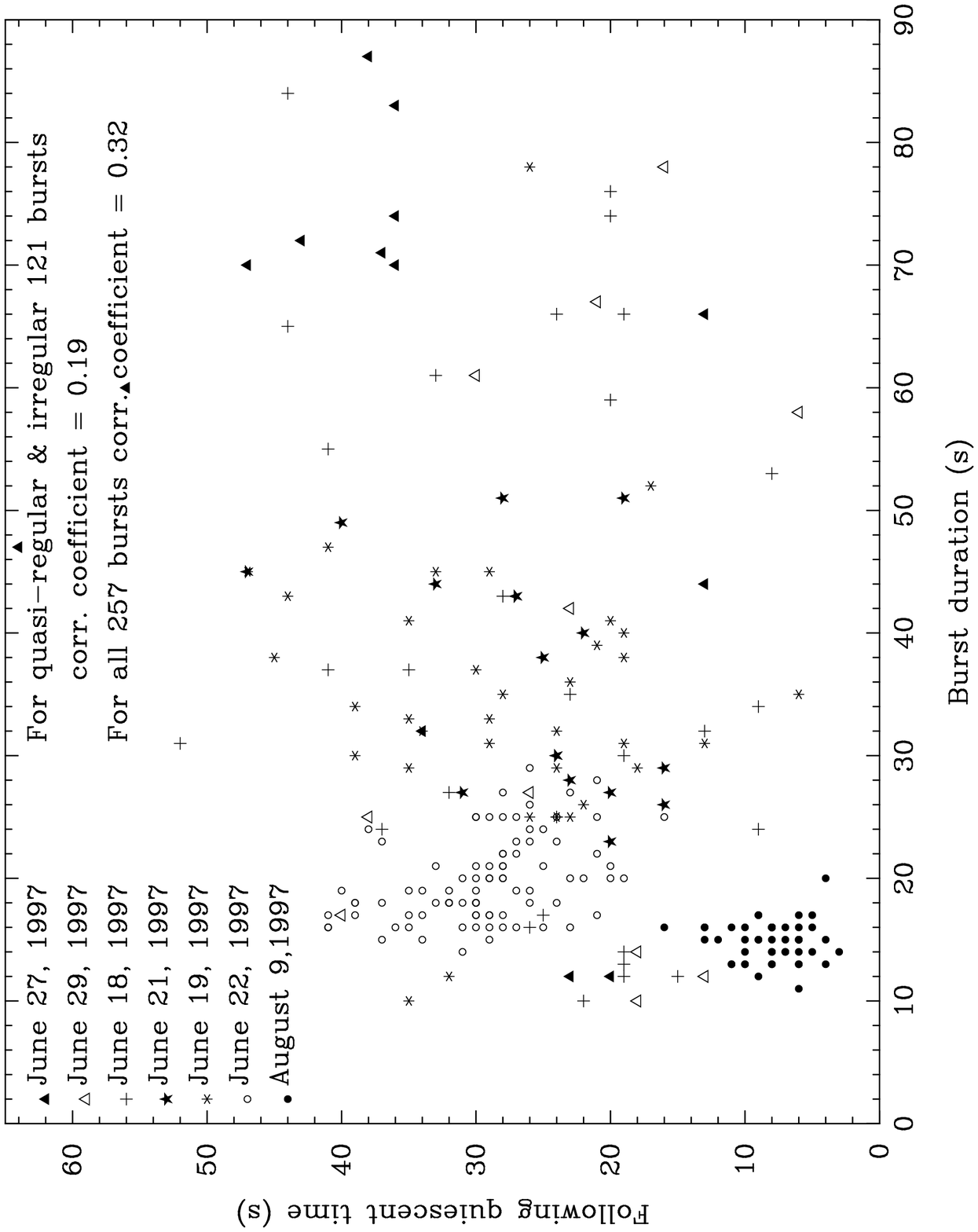]{Correlation between following quiescent time and the burst duration
(as defined in the text) for the same data as in Figure 5. \label{fig7}}

\figcaption[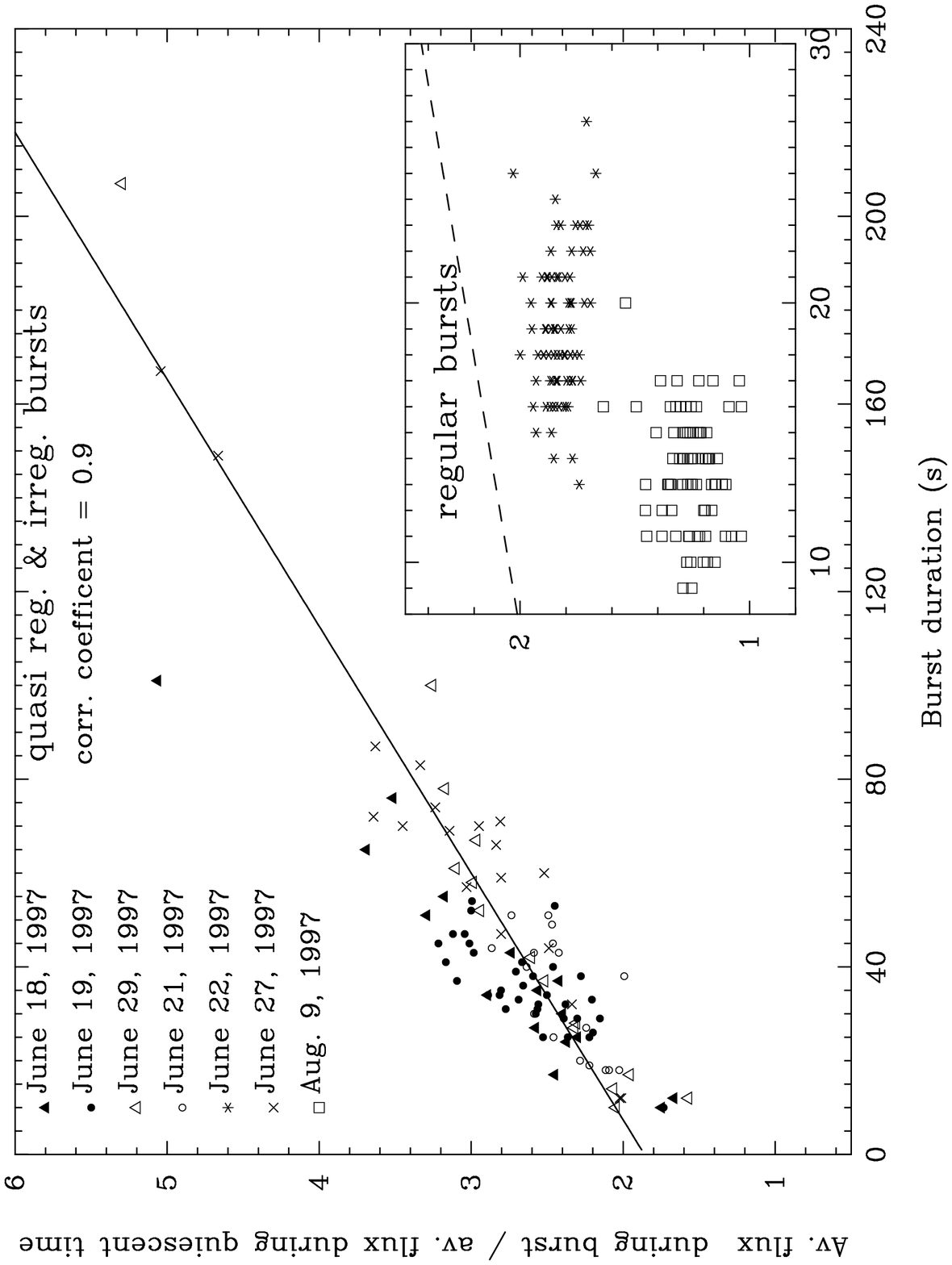]{The ratio of the average flux during burst to the average flux during quiescent time as a function of the burst duration for the quasi-regular and irregular bursts. The solid line is the least square fit to the data. In the inset, results for both the type of regular bursts are shown. The dotted line is the fit for the quasi-regular and irregular bursts shown for comparison. \label{fig8}}

\figcaption[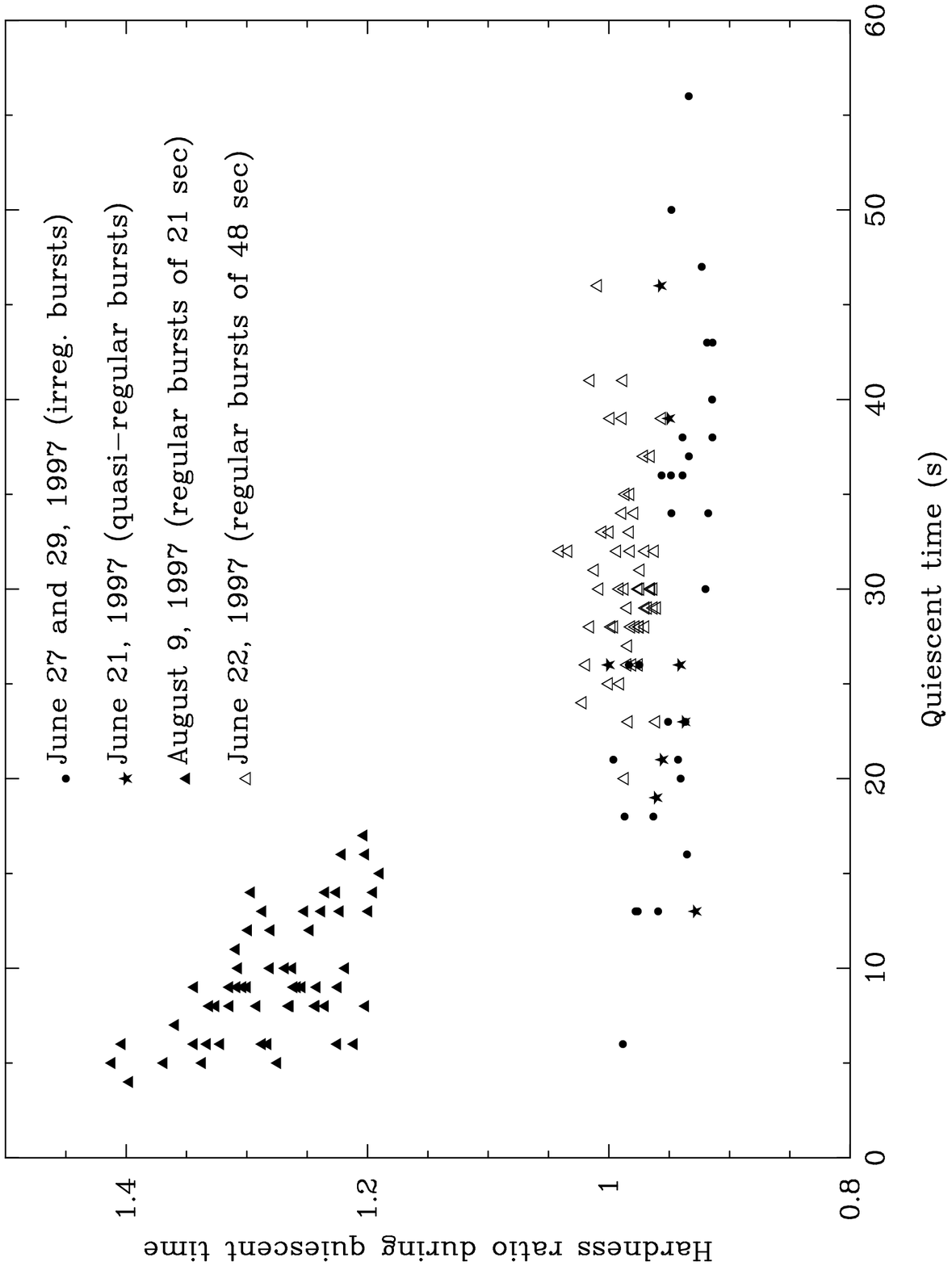]{The hardness ratio during  the quiescent phase vs the quiescent time for 
all the type of observed bursts from GRS~1915+105. \label{fig9}}

\figcaption[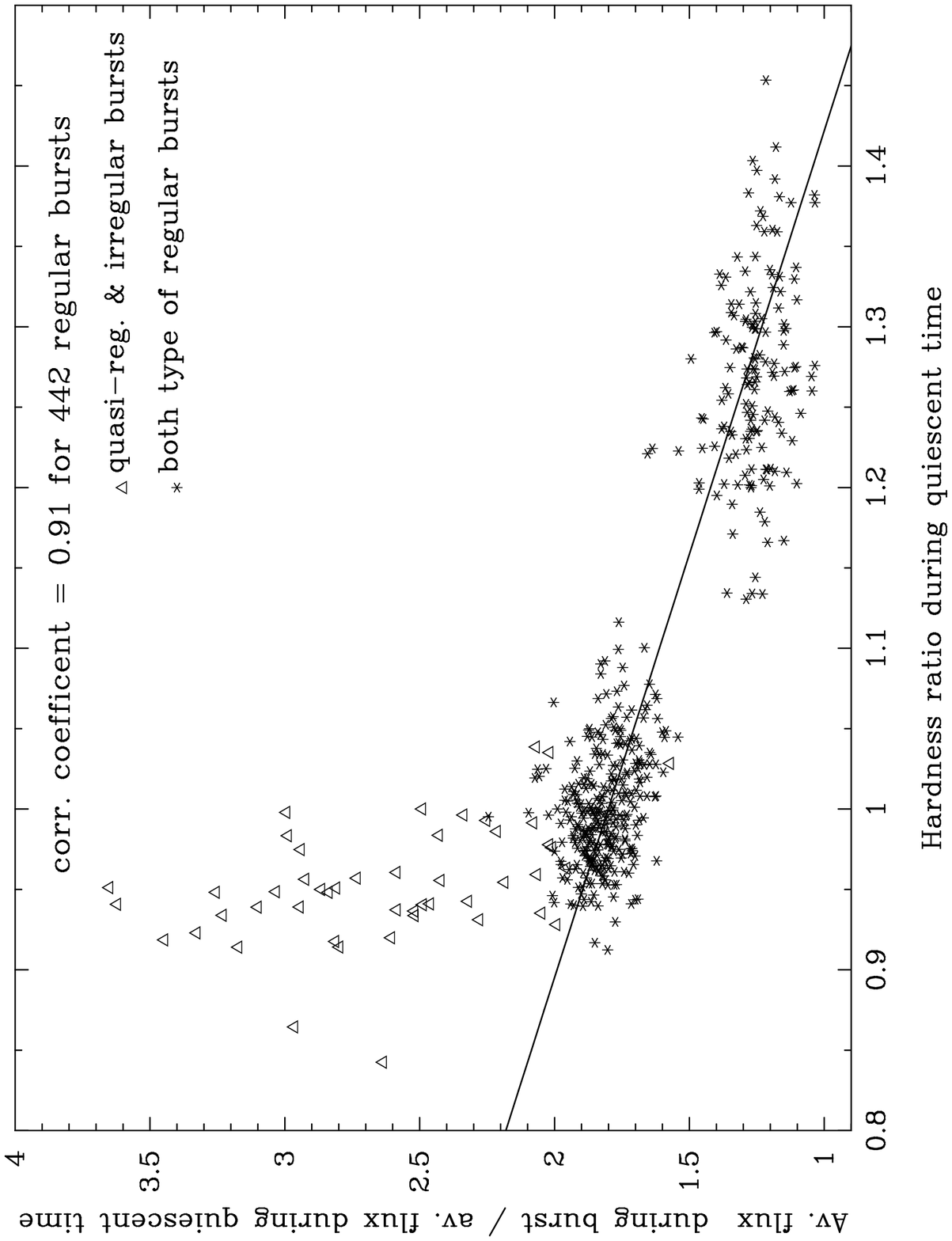]{The ratio of the average flux during burst to the average flux during the quiescent time as a function of the hardness ratio during the quiescent time for all the type of observed  bursts. The least square fit to  both the type of regular bursts is shown by a solid line.  \label{fig10}}

\figcaption[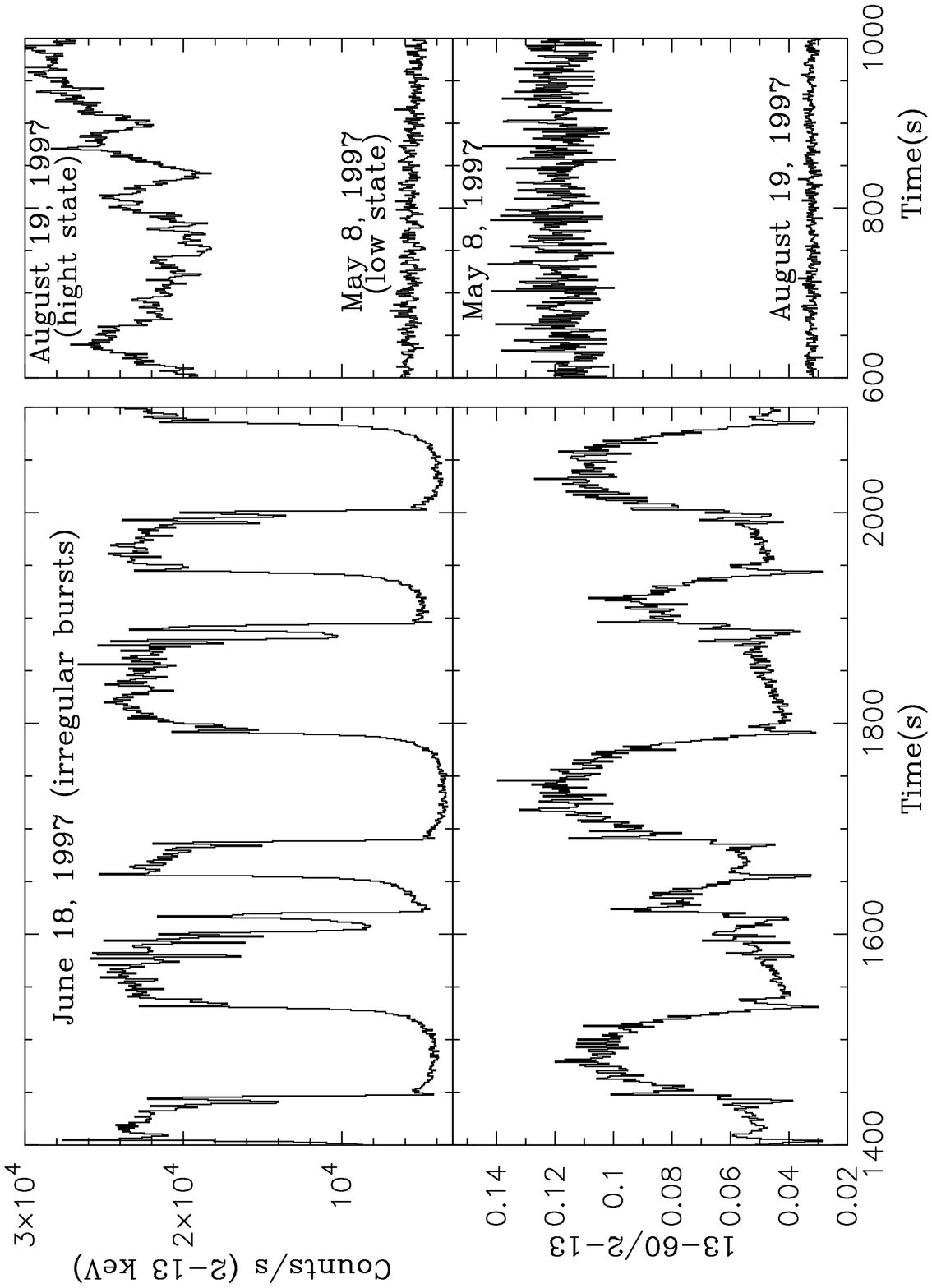]{Plot of 2$-$13 keV flux vs time is shown in left top panel (irregular bursts) and the ratio of  13$-$60 keV flux to 2$-$13 keV flux is plotted as a function of time in the left bottom panel from  RXTE/PCA data of 1997 June 18. In the right top panel, 2$-$13 keV  flux is plotted from RXTE/PCA data of 
1997 August 19 and May 8 when source was in  high and low states respectively. 
 Respective ratios of 13$-$60 keV flux to 2$-$13 keV flux are plotted in the right bottom panel. \label{fig11}} 

\figcaption[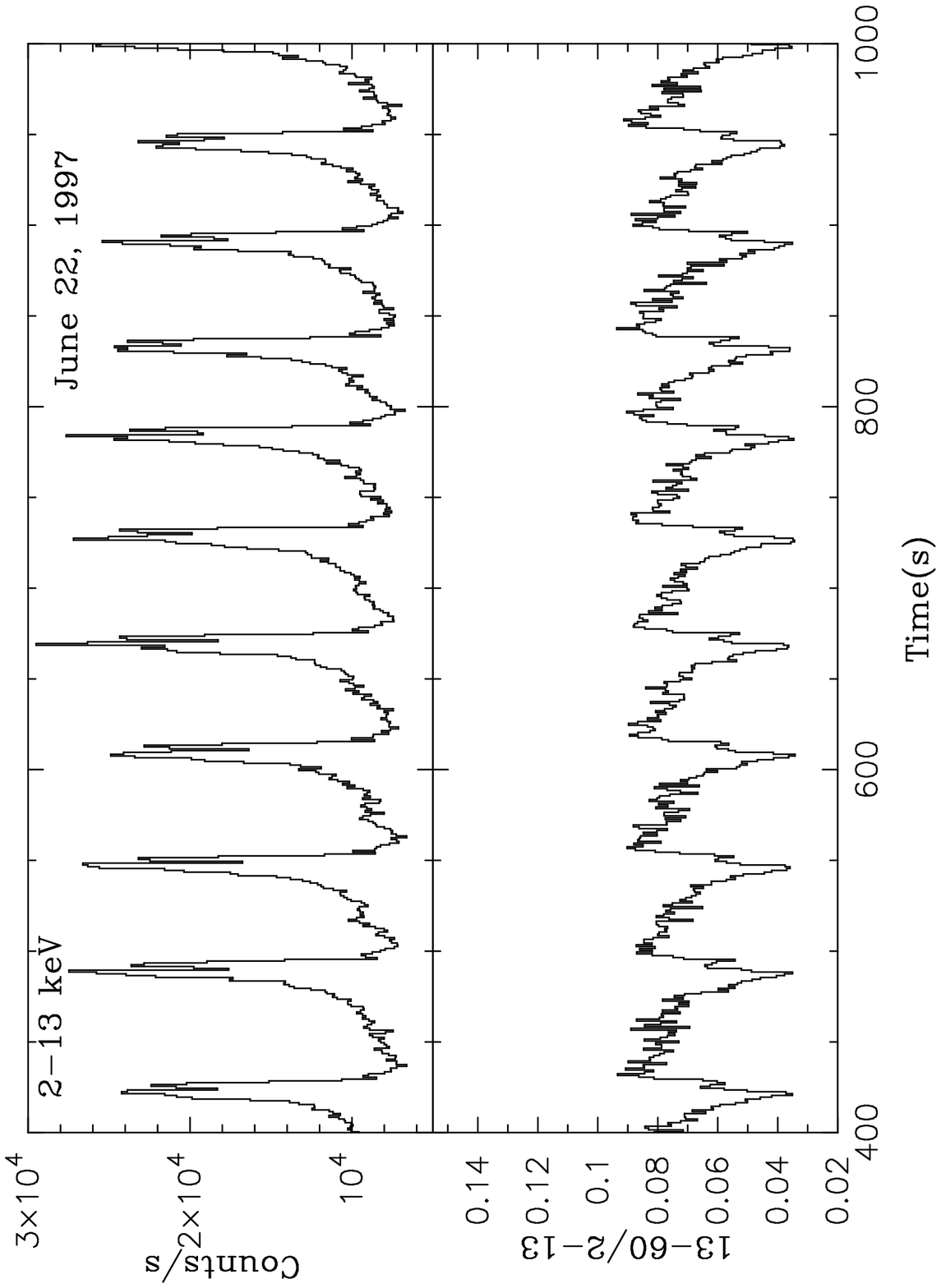]{Plot of 2$-$13 keV flux vs time is shown in the top panel (regular bursts) and the ratio of 13$-$60 keV flux to 2$-$13 keV flux is plotted as a function of time in the bottom panel from  RXTE/PCA data of 1997 June 22. \label{fig12}}

\setcounter{figure}{0}
\clearpage
\begin{figure}
\plotone{fig1.eps}
\end{figure}
\clearpage
\begin{figure}
\plotone{fig2.eps}
\end{figure}
\clearpage
\begin{figure}
\vbox to6in{\rule{0pt}{6in}}
\includegraphics{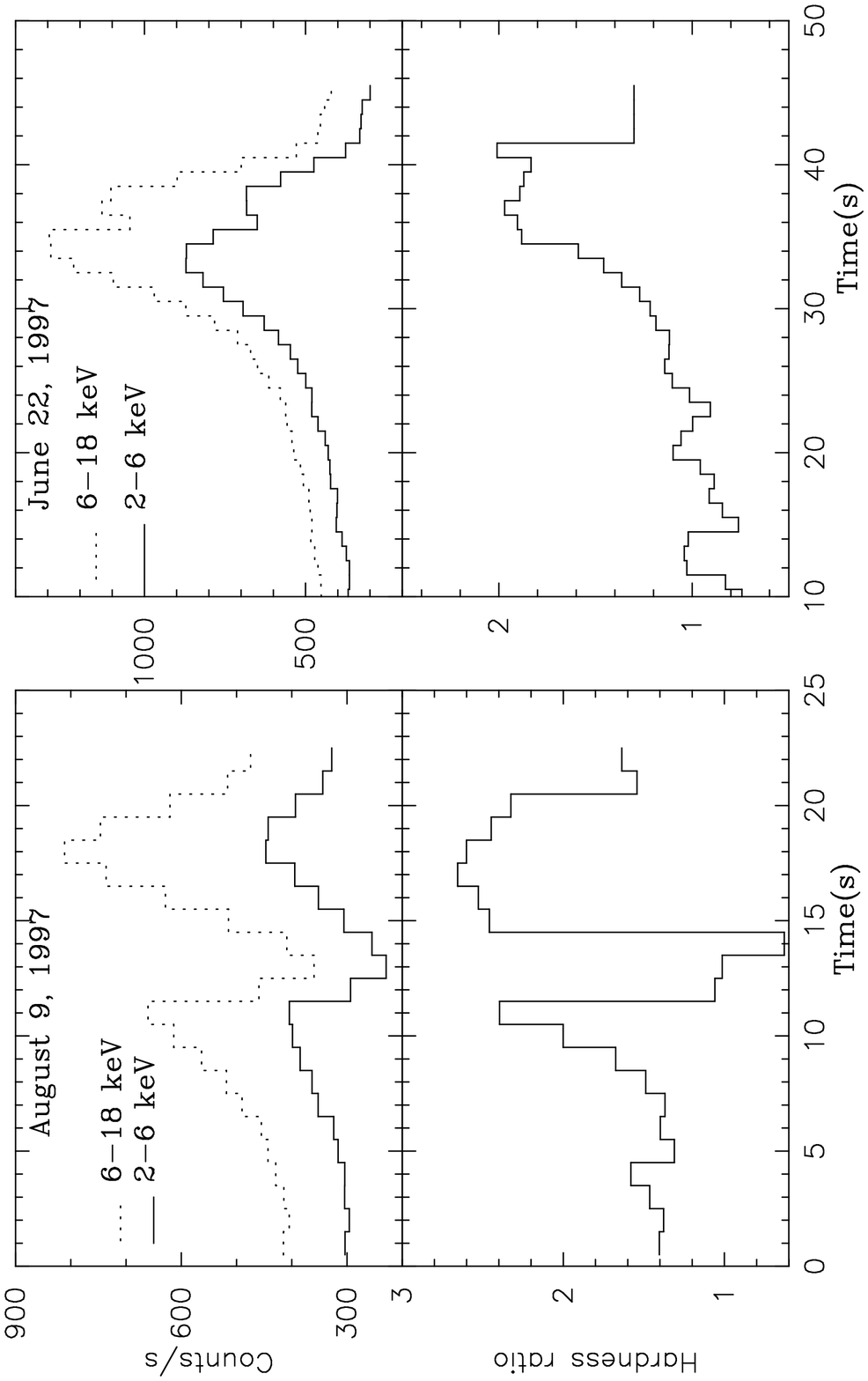}
\end{figure}
\clearpage
\begin{figure}
\vbox to6in{\rule{0pt}{6in}}
\includegraphics{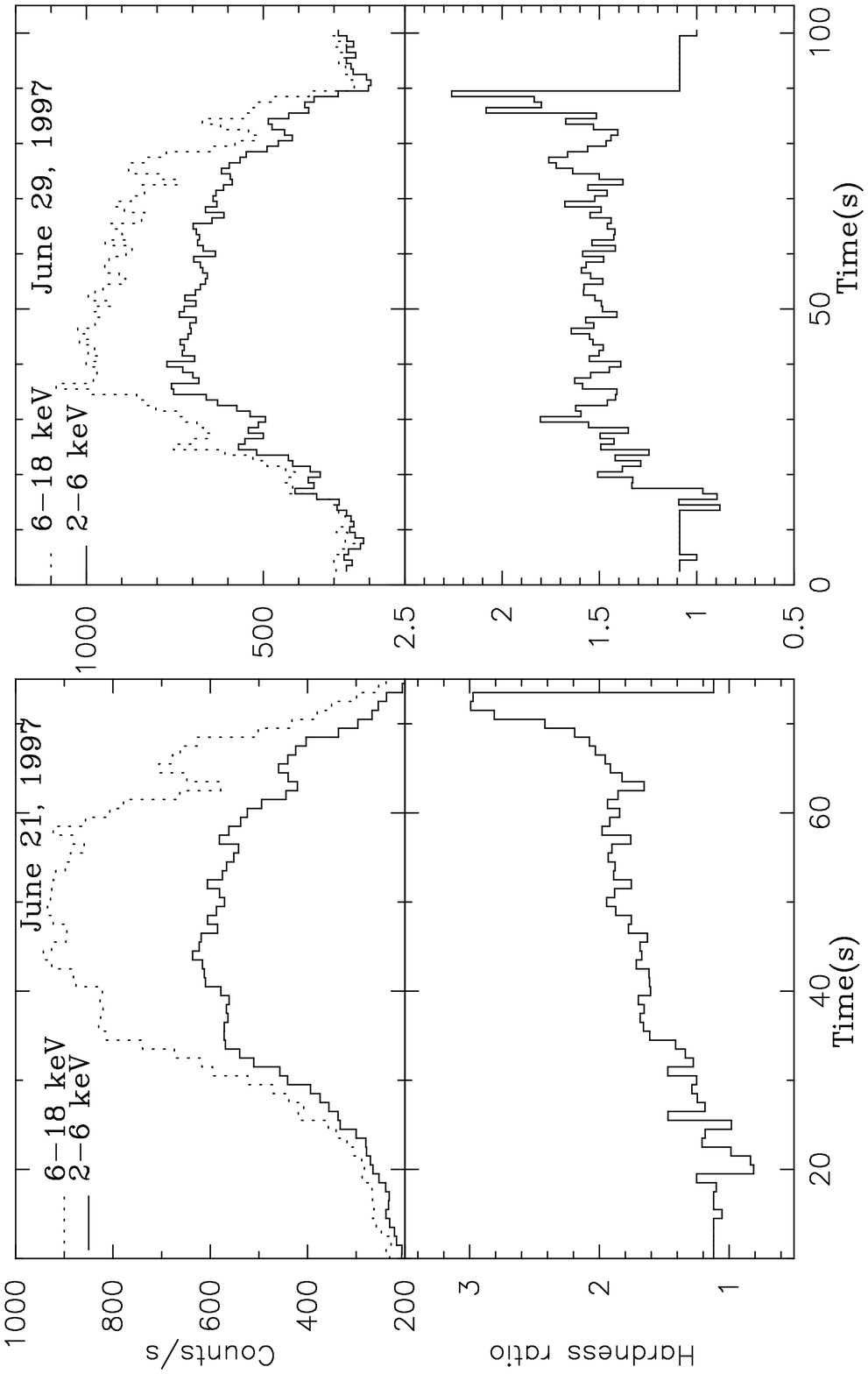}
\end{figure}

\clearpage
\begin{figure}
\plotone{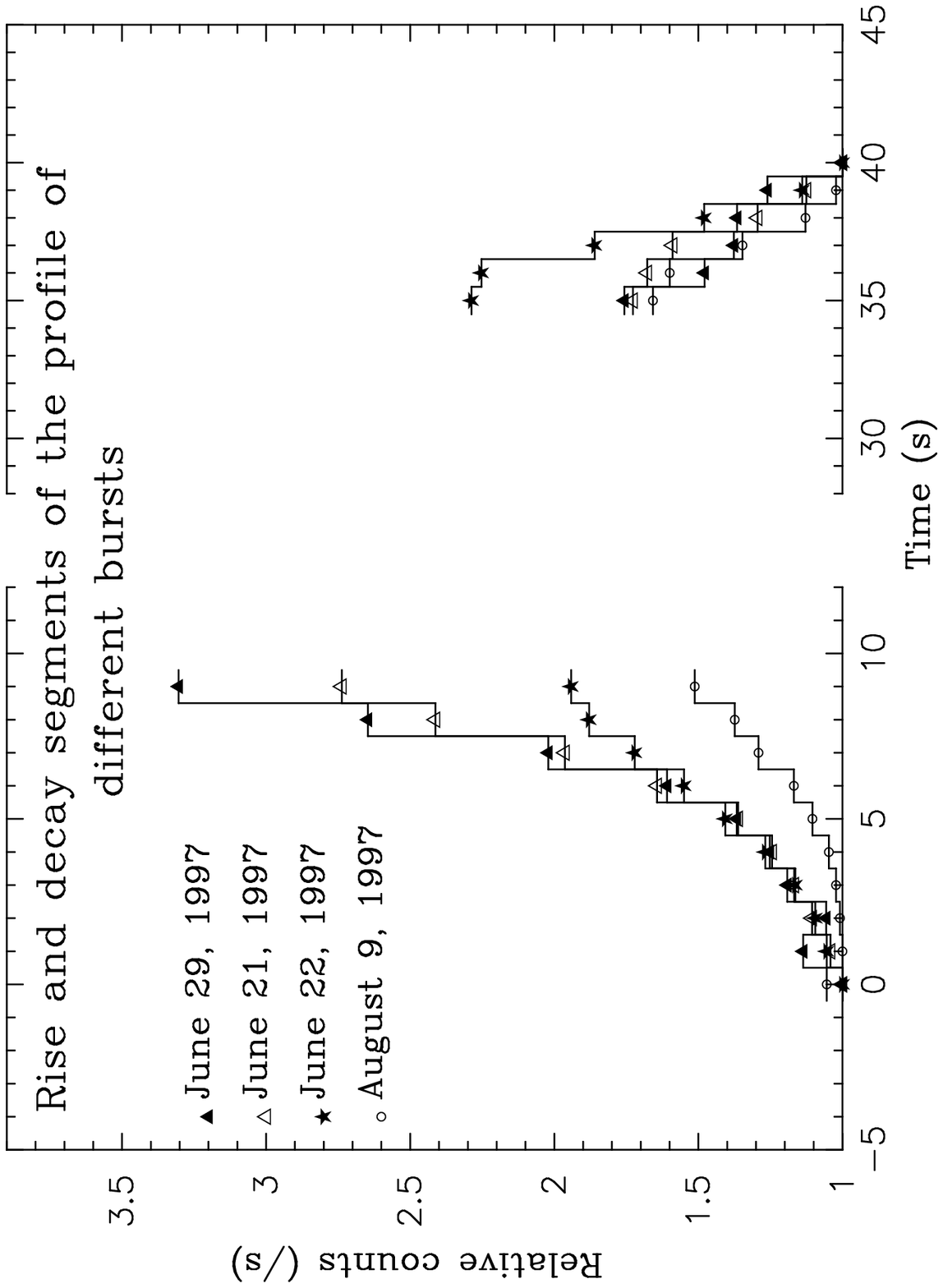}
\end{figure}

\clearpage
\begin{figure}
\plotone{fig5.eps}
\end{figure}

\clearpage
\begin{figure}
\plotone{fig6.eps}
\end{figure}

\clearpage
\begin{figure}
\plotone{fig7.eps}
\end{figure}

\clearpage
\begin{figure}
\plotone{fig8.eps}
\end{figure}

\clearpage
\begin{figure}
\plotone{fig9.eps}
\end{figure}

\clearpage
\begin{figure}
\plotone{fig10.eps}
\end{figure}

\clearpage
\begin{figure}
\plotone{fig11.eps}
\end{figure}

\clearpage
\begin{figure}
\plotone{fig12.eps}
\end{figure}
\end{document}